\documentclass[aps,pra,twocolumn,showpacs,floatfix,nofootinbib,groupedaddress,superscriptaddress]{revtex4}
\usepackage{amsfonts,amssymb,amsmath,bm,bbm,graphics,graphicx,times,color,multirow,ulem}
\newcommand{\ignore}[1]{}
\begin{document}

\title{Quantum Process Estimation via Generic Two-Body Correlations}
\author{M. Mohseni}
\affiliation{Research Laboratory of Electronics, Massachusetts
Institute of Technology, 77 Massachusetts Avenue, Cambridge MA
02139, USA}
\affiliation{Department of Chemistry and Chemical
Biology, Harvard University, 12 Oxford St., Cambridge, MA 02138,
USA}
\author{A. T. Rezakhani}
\affiliation{Department of Chemistry, Center for Quantum Information
Science and Technology, University of Southern
California, Los Angeles, CA 90089, USA}
\author{J. T. Barreiro}
\affiliation{Department of Physics, University of Illinois at
Urbana-Champaign, Urbana, IL 61801, USA} \affiliation{Institut f\"{u}r Experimentalphysik, Universit\"{a}t Innsbruck, Technikerstrasse 25/4,
A-6020 Innsbruck, Austria}
\author{P. G. Kwiat}
\affiliation{Department of Physics, University of Illinois at
Urbana-Champaign, Urbana, IL 61801, USA}
\author{A. Aspuru-Guzik}
\affiliation{Department of Chemistry and Chemical Biology, Harvard
University, 12 Oxford St., Cambridge, MA 02138, USA}

\begin{abstract}
Performance of quantum process estimation is naturally limited to
fundamental, random, and systematic imperfections in preparations
and measurements. These imperfections may lead to considerable
errors in the process reconstruction due to the fact that standard
data analysis techniques presume ideal devices. Here, by utilizing
generic auxiliary quantum or classical correlations, we provide a
framework for estimation of quantum dynamics via a single
measurement apparatus. By construction, this approach can be applied
to quantum tomography schemes with calibrated faulty state
generators and analyzers. Specifically, we present a generalization
of \textquotedblleft Direct Characterization of Quantum Dynamics"
[M. Mohseni and D. A. Lidar, Phys. Rev. Lett.  \textbf{97}, 170501 (2006)] with an
imperfect Bell-state analyzer. We demonstrate that, for several
physically relevant noisy preparations and measurements, only
classical correlations and small data processing overhead are
sufficient to accomplish the full system identification.
Furthermore, we provide the optimal input states for which the error
amplification due to inversion on the measurement data is minimal.
\end{abstract}

\maketitle


\section{Introduction}

Quantum measurement theory imposes fundamental limitations on the
information extractable from a quantum system. Although the
evolution of quantum systems can be described deterministically, the
measurement operation always leads to nondeterministic outcomes. In
order to obtain a desired accuracy, measurement of a particular
observable needs to be repeated over an ensemble of identical
quantum systems. In addition, for systems with many degrees of
freedom, one usually needs to measure a set of non-commuting
observables corresponding to independent parameters of the system.
Characterization of state or dynamics of a quantum system can be
achieved by a family of methods known as quantum tomography
\cite{qt,nielsen-book}. In particular, quantum process tomography
provides a general experimental procedure for estimating dynamics of
a system which has an unknown interaction with its embedding
environment for discrete or continuous degrees of freedom \cite
{nielsen-book,d'ariano-aapt,dcqd1,kosut08,alex08}. In these methods,
the full information is obtained by a complete set of experimental
settings associated with the set of required input states and
non-commuting measurements. In recent developments
\cite{d'ariano-aapt,altepeter-aapt,dcqd1,dcqd2,WangExDCQD07,bendersky-paz},
it has been demonstrated that the minimal number of required
experimental settings can indeed be substantially reduced by using
degrees of freedom of auxiliary quantum systems correlated with the
system of interest.

Generally it is possible to completely characterize a quantum device
with a single experimental setting. A correlated input state of the
combined system and an ancilla is subjected to the unknown process,
and a generalized measurement or Positive Operator Valued Measure
(POVM) is performed at the output \cite{Dariano02,dcqd3}. However,
in order to realize such a generalized measurement one needs to
effectively generate many-body interactions \cite{dcqd3} which are
not naturally available and/or controllable. Quantum simulation of
such many-body interactions is in principle possible, but generally
requires an exponentially large number of single- and two-body
interactions with respect to system's degrees of freedom. An
alternative method that circumvents the requirement for many-body
interactions, yet allows simultaneous non-commuting observables
through a single measurement setting, is known as Direct
Characterization of Quantum Dynamics (DCQD) \cite{dcqd1,dcqd2}. The
construction of the full information about the dynamical process is
then possible via preparation of a set of mutually unbiased
entangled input states over a subspace of the total Hilbert space of
the principle system and an ancilla \cite{dcqd2}. The DCQD approach
was originally developed with the assumptions of ideal (i.e.,
error-free) quantum state preparation, measurement, and ancilla
channels. However, in a realistic estimation process, due to
decoherence, limited preparation/measurement accuracies, or other
imperfections, certain errors may occur hindering the overall
procedure.

In this work, we introduce an experimental procedure for using
generic two-body interactions to perform quantum process estimation
on a subsystem of interest. We employ this approach to generalize
the DCQD
scheme
to the cases in which the preparations and
measurements are realized with known systematic faulty operations.
We demonstrate that in some specific, but physically motivated,
noise models, such as the generalized depolarizing channels, only
classical correlation between the system and ancilla suffice.
Moreover, for these situations the data processing overhead is fairly small in
comparison to the ideal DCQD. Given a noise model,
one can find the optimal input states by minimizing the errors
incurred through the inversion of experimental data. Thus, we
provide the optimal input states for reducing the inversion errors
in the noiseless DCQD scheme.

The structure of the paper is as follows. In Sec.~\ref{general}, we
set the framework for process tomography schemes where we have faulty --- rather than ideal --- faulty Bell-state analyzers, emphasizing the DCQD approach. Next, in
Sec.~\ref{models}, we demonstrate the applicability of our framework
through some simple, yet important examples of noise models. We
conclude by summarizing the paper in Sec.~\ref{summary}.

\section{Characterization of quantum processes with a faulty Bell-state analyzer}
\label{general}

Let us consider a given quantum system composed of two correlated
physical subsystems $A$ and $B.$ For a time duration
$\Delta t$
the two
subsystems are decoupled, thus experiencing different quantum
processes, and then they interact with each other again.  The task
is to characterize the unknown quantum process acting on the
subsystem of interest $A$, assuming we have prior knowledge about the
dynamics of subsystem $B$ plus their initial and final correlations.
Another similar scenario can also be envisioned. Given two controllable
quantum systems $A$ and $B$ that are made to sufficiently interact
before and after a time duration
$\Delta t$,
we wish to estimate
the unknown dynamics acting on system $A$ for such time interval,
assuming the dynamics of the ancilla system $B$ and the interaction
between two systems is known with a certain accuracy.

Much progress has been made in creating and characterizing two-body
correlations in a variety of physical systems and interactions,
including nuclear magnetic resonance (NMR) systems interacting through an Ising Hamiltonian
together with refocusing or dynamical decoupling techniques
\cite{NMR-}, atoms/molecules in cavity quantum electrodynamics (QED) \cite{QED-}, trapped ions
interacting via the Jaynes-Cummings Hamiltonian driven by laser
pulses and vibrational degrees of freedom \cite{ion-trap}, and
photons correlated in one or many degrees of freedom, e.g.,
generated by parametric-down conversion \cite{optics} or four-wave
mixing \cite{four-wave}.
Other approaches include spin-coupled quantum dots
\cite{QuantumDots}, superconducting qubits \cite{Superconducting}
controlled by external electric and/or magnetic fields, and
chromophoric complexes coupled through Forster/Dexter interactions
and monitored or controlled via ultra-fast nonlinear spectroscopy
\cite{Chromophor}. However, in almost all of these systems, the
entangled Bell-state preparations and measurements, which
generically are the basic building blocks of quantum information
processing, hardly achieve high fidelities; at the very least they
are certain to be imperfect at some level, and this will limit their
use for tomography.  Our goal is to determine the optimal states and
measurement strategy that will minimize the deleterious effects of
the nonidealities --- assumed known --- on process tomography.

\begin{figure}[tp]
\includegraphics[width=8cm]{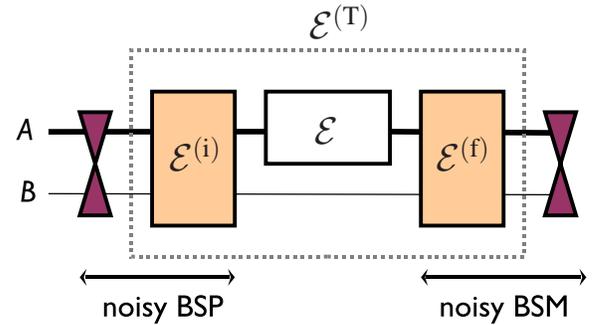}
\begin{picture}(0,10)
\put(-113,70){\Large$\mathcal{E}$}
\put(-115,110){\Large$\mathcal{E}^{(\text{T})}$}
\put(-168,60){\Large$\mathcal{E}^{(\text{i})}$}
\put(-67,60){\Large$\mathcal{E}^{(\text{f})}$}
\end{picture}
\caption{(color online) Schematic of a faulty DCQD, with imperfect
or noisy Bell-state preparation (BSP) and Bell-state measurement
(BSM). }
\label{gdcqd-fig}
\end{figure}

We consider the cases where we can
simulate initial or final two-body correlations in the above
schemes by performing an ideal (generalized) Bell-state preparation
(or measurement) followed by a \textit{known} faulty
completely-positive (CP) quantum map acting on the entire systems involved.
It should be noted that not every CP maps can be written as
concatenation of two other CP maps. In other words, there exist CP
maps that are ``indivisible'', in the sense that, for such a map
$\mathcal{T}$, there do not exist CP maps $\mathcal{T}_1$ and
$\mathcal{T}_2$ such that $\mathcal{T}=\mathcal{T}_2 \mathcal{T}_1$,
where neither $\mathcal{T}_1$ or $\mathcal{T}_2$ are unitary
\cite{wolf}. Nonetheless, all full-rank CP maps --- in the sense of
the Kraus representation \cite{nielsen-book} --- are divisible.

Here,
we include
quantum maps acting on
system $B$ 
in the preparation or measurement maps.  This approach naturally provides a
generalization of the DCQD scheme to the cases of faulty
preparations, measurement and ancilla channels where the noise is
already known --- see Fig.~\ref{gdcqd-fig}. For simplicity, in this
work we concentrate only on one-qubit systems and the DCQD scheme
(summarized
in
Table~\ref{dcqd-tab}). However, generalization of the
framework is straightforward for other process estimation schemes
and for DCQD on qudit systems with $d$ being a power of a prime
(according to the construction of Ref. ~\cite{dcqd2}).

\begingroup
\squeezetable
\begin{table}[bp]
\begin{ruledtabular}
\caption{The ideal direct characterization of single-qubit
$\bm{\chi}$. Here
$|\Phi^{+}_{\alpha}\rangle=\alpha|00\rangle+\beta|11\rangle$,
$|\Phi^{+}_{\alpha}\rangle_{x(y)}=\alpha|++\rangle_{x(y)}+\beta|--\rangle_{x(y)}$
where $|\alpha |\neq |\beta |\neq 0$ and $\mathrm{Im}(\bar{\alpha}
\beta)\neq 0$, and $\{|0\rangle,|1\rangle \}$, $\{|\pm\rangle_x \}$,
$\{|\pm\rangle_y\}$ are eigenstates of the Pauli operators
$\sigma_z$, $\sigma_x$, and $\sigma_{y}$. $P_{\Phi^+}$ is the
projector on the Bell state $|\Phi^+\rangle$, and similarly for the
other projectors. See Refs.~\cite{dcqd1,dcqd3}. Determination of
optimal values of
$\alpha$ and $\beta$
is discussed in the text.}
\begin{tabular}{l|c|c}
input state & Bell-state measurement & output $\chi_{mn}$\\
 \colrule
$|\Phi^+\rangle$  &$P_{\Psi^{\pm}},P_{\Phi^{\pm}}$ &
$\chi_{00},\chi_{11},\chi_{22},\chi_{33}$\\
$|\Phi^+_{\alpha}\rangle$ & $P_{\Phi^{+}}
\pm P_{\Phi^{-}}, P_{\Psi^{+}}\pm P_{\Psi^{-}}$ & $\chi_{03},\chi_{12}$\\
$|\Phi^+_{\alpha}\rangle_x$ & $P_{\Phi^{+}}\pm P_{\Psi^{+}},
P_{\Phi^{-}}\pm P_{\Psi^{-}}$ & $\chi_{01},\chi_{23}$\\
$|\Phi^+_{\alpha}\rangle_y$  & $P_{\Phi^{+}}\pm P_{\Psi^{-}},
P_{\Phi^{-}}\pm P_{\Psi^{+}}$ & $\chi_{02},\chi_{13}$\\
\end{tabular}\label{dcqd-tab}
\end{ruledtabular}
\end{table}
\endgroup

Let us consider the qubit of interest $A$ and the
ancillary qubit
$B$
prepared in the maximally entangled state $|\Phi ^{+}\rangle
_{AB}=(|00\rangle +|11\rangle )_{AB}/\sqrt{2}$.
We first apply a \textit{known} quantum error map $\mathcal{E}^{(\text{i%
})}$ to $A$ and $B$: $\mathcal{E}^{(\text{i})}(\rho
)=\sum_{pqrs}\chi _{pqrs}^{(\text{i})}\sigma _{p}^{A}\sigma
_{q}^{B}\rho \sigma _{r}^{B}\sigma _{s}^{A}$, where $\rho =|\Phi
^{+}\rangle \left\langle \Phi ^{+}\right\vert $ and $\left\{ \sigma
_{0}\equiv \openone,\sigma _{1},\sigma _{2},\sigma _{3}\right\} $
are the identity and Pauli operator for a single qubit. Next, we
apply the
unknown quantum map $\mathcal{E}$ to qubit A; this is what we are trying to determine: $\mathcal{E}[\mathcal{E%
}^{(\text{i})}(\rho )]=\sum_{mn}\chi _{mn}\sigma
_{m}^{A}(\sum_{pqrs}\chi _{pqrs}^{(\text{i})}\sigma _{p}^{A}\sigma
_{q}^{B}\rho \sigma _{r}^{B}\sigma _{s}^{A})\sigma _{n}^{A}.$
Finally, we apply a \textit{known} quantum error map before the
Bell-state measurement. Note that in this approach any error on the
ancilla channel can be absorbed into either $\mathcal{E}^{(\text{f})}$ or $\mathcal{E%
}^{(\text{i})}$. The total map acting on the combined system $AB$ is then $%
\mathcal{E}^{(\text{T})}=\mathcal{E}^{(\text{f})}\circ
\mathcal{E}\circ \mathcal{E}^{(\text{i})}$, given by
\begin{eqnarray*}
\mathcal{E}^{(\text{T})}(\rho ) &=&\sum_{mnpp^{\prime }qq^{\prime
}rr^{\prime }ss^{\prime }}\chi _{mn}\chi _{pqrs}^{(\text{i})}\chi
_{p^{\prime }q^{\prime }r^{\prime }s^{\prime }}^{(\text{f})}\times \\
&&\sigma _{p^{\prime }}^{A}\sigma _{q^{\prime }}^{B}\sigma
_{m}^{A}\sigma _{p}^{A}\sigma _{q}^{B}\rho \sigma _{r}^{B}\sigma
_{s}^{A}\sigma _{n}^{A}\sigma _{r^{\prime }}^{B}\sigma _{s^{\prime
}}^{A},
\end{eqnarray*}%
where the parameters $\chi _{pqrs}^{(\text{i})}$ and $\chi
_{p^{\prime }q^{\prime }r^{\prime }s^{\prime }}^{(\text{f})}$ are
known (from calibration of the operational/systematic errors in the
preparation and measurement devices). By defining $\omega
_{mnp^{\prime }s^{\prime }}=(-1)^{(\delta _{m0}-1)(\delta
_{p^{\prime }0}-1)\delta _{mp^{\prime
}}+(\delta _{n0}-1)(\delta _{s^{\prime }0}-1)\delta _{ns^{\prime }}}$ and $%
\widetilde{\rho }_{mn}=\sum_{pp^{\prime }qq^{\prime }rr^{\prime
}ss^{\prime }}\omega _{mnp^{\prime }s^{\prime }}\chi
_{pqrs}^{(\text{i})}\chi _{p^{\prime }q^{\prime }r^{\prime
}s^{\prime }}^{(\text{f})}\sigma _{p^{\prime }}^{A}\sigma
_{q^{\prime }}^{B}\sigma _{p}^{A}\sigma _{q}^{B}\rho \sigma
_{r}^{B}\sigma _{s}^{A}\sigma _{r^{\prime }}^{B}\sigma _{s^{\prime
}}^{A}$, we have
\begin{eqnarray*}
& \mathcal{E}^{(\text{T})}(\rho )=\sum_{mn}\chi _{mn}\sigma _{m}^{A}\widetilde{%
\rho }_{mn}\sigma _{n}^{A}.
\end{eqnarray*}%
By construction, the parameters $\chi _{pqrs}^{(\text{i})}$ and
$\chi _{p^{\prime }q^{\prime }r^{\prime }s^{\prime }}^{(\text{f})}$
are all \textit{a priori} known, as are the matrices $\widetilde{\rho }_{mn}$ which are functions of $%
\bm{\chi}^{(\text{i})}$, $\bm{\chi}^{(\text{f})}$, and the initial state $%
\rho $. Therefore, in order to develop a generalized DCQD scheme for
the systems with faulty Bell-state preparation (BSP) and measurement
(BSM), we need to do it for a set of modified (input) states rather
than a pure Bell-state type input. Expanding $\widetilde{\rho
}_{mn}$ in the Bell basis yields
\begin{eqnarray*}
& \widetilde{\rho }_{mn}=\sum_{kk'}\lambda _{mn}^{kk'}P^{kk'},
\end{eqnarray*}%
where $\lambda _{mn}^{kk'}=\text{Tr}[P^{kk'}\widetilde{\rho }%
_{mn}]$, $P^{kk'}=\left\vert B^{k}\right\rangle
\left\langle B^{k'}\right\vert$, and $|B^{k}\rangle $ for $k=0,1,2,3$ corresponds to the Bell-states
$\left\vert \Phi ^{+}\right\rangle $, $\left\vert \Psi
^{+}\right\rangle $, $\left\vert \Psi ^{-}\right\rangle $, and
$\left\vert \Phi ^{-}\right\rangle $,
respectively, where $|\Phi ^{\pm }\rangle =(|00\rangle \pm |11\rangle )/%
\sqrt{2},~|\Psi ^{\pm }\rangle =(|01\rangle \pm |10\rangle
)/\sqrt{2}$. (Henceforth throughout this manuscript, superscripts
refer to the Bell-state basis and subscripts refer to the Pauli operator
basis.) The $\lambda _{mn}^{kk^{\prime }}$s are known functions of
$\omega _{mnp^{\prime
}s^{\prime }}$, $\bm{\chi}^{(\text{f})}$, $\bm{\chi}^{(\text{i})}$, and $%
\rho $. Therefore, the overall output state can be rewritten as follows:
\begin{eqnarray*}
& \mathcal{E}^{(\text{T})}(\rho )=\sum_{kk^{\prime }mn}\lambda
_{mn}^{kk^{\prime }}\chi _{mn}\sigma _{m}^{A}P^{kk^{\prime }}\sigma
_{n}^{A}.
\end{eqnarray*}

We now apply the standard DCQD data analysis to estimate the matrix
elements of $\bm{\chi}^{(\text{T})}$ (representing
$\mathcal{E}^{(\text{T})}$). \ After performing a BSM, i.e.,
measuring $\{P^{jj}\}_{j=0}^{3}$
on this state,
we obtain the Bell-state $|B^{j}\rangle $ with probability
\begin{eqnarray*}
& \mathrm{Tr}[P^{jj}\mathcal{E}^{(\text{T})}(\rho
)] =\sum_{kk^{\prime }mn}\Lambda _{kk^{\prime },mn}^{(j)}\chi _{mn},
\end{eqnarray*}%
where $\Lambda _{kk',mn}^{(j)}=\lambda _{mn}^{kk'}\mathrm{%
  Tr}[P^{jj}\sigma_{m}^{A}P^{kk'}\sigma _{n}^{A}]$. Although this
expression can be made more compact by using Pauli identities, the
current form is convenient for our purposes.

\ignore{ \footnote{This form can be made
more compact by using the identity $\sigma
_{m}^{A}P^{kk'}\sigma _{n}^{A}=P^{ii'}$ for some $i$ and $%
i'$ which are (known) functions of $m$, $n$, $k$, and $k'$. More explicitly,
we take use of (i) $\sigma _{k}^{A}|\Phi^{+}\rangle
=(-i)^{\alpha _{k}}|B^{k}\rangle $, where $\alpha _{k}=0$ for
$k=0,1,3$, and $\alpha_{2}=1$, (ii) $\sigma _{m}^{A}P^{kk'}\sigma _{n}^{A}
=i^{\alpha _{k} + \alpha _{k'}}\sigma_{m}^{A}\sigma_{k}^{A}|\Phi ^{+}\rangle \langle \Phi ^{+}|\sigma
_{k'}^{A}\sigma_{n}^{A}$, and (iii) $\sigma _{m}\sigma _{n}
=\delta _{mn}\sigma _{0}+\sum_{l}\widetilde{\epsilon }_{mnl}\sigma _{l}$, where
\begin{equation*}
\widetilde{\epsilon }_{mnl}=%
\begin{cases}
i\epsilon _{mnl}:~~m,n,l=1,2,3 \\
\delta _{nl}:~~m=0\neq n \\
\delta _{ml}:~~n=0\neq m \\
0:~~\text{a pair of indices are zero}%
\end{cases},
\end{equation*}
to find the compact functional forms of $i$ and $i'$.} }

A similar set of equations for the standard DCQD inputs $\{\rho
^{(i)}\}_{i=0}^{3}$ can also be written.  We represent all of these
equations in a compact vector form as
\begin{eqnarray}
|\bm{\chi}^{(\text{T})}) =\mathbf{\Lambda}|\bm{\chi}),
\label{vector-eq}
\end{eqnarray}
where the
$\bm{\Lambda}(\bm{\chi}^{(\text{i})},\bm{\chi}^{(\text{f})},\{\rho
^{(i)}\})$ matrix contains full information about all faulty experimental conditions. Given $\bm{\chi}^{(%
\text{i})}$, $\bm{\chi}^{(\text{f})}$, and the standard DCQD input
set $\{\rho ^{(i)}\} $, one can calculate the
$\bm{\Lambda}$ matrix. The standard DCQD experimental data (analysis) will also determine $|\bm{\chi}^{(%
\text{T})})$. Now, if the $\bm{\Lambda}$ matrix is invertible, from
Eq.~(\ref{vector-eq}) one can obtain $\bm{\chi}$ by
inversion: $|\bm{\chi}) =\bm{\Lambda}^{-1}|\bm{\chi}^{(\text{T}%
)})$. The invertibility of the $\bm{\Lambda}$ matrix, namely $\det %
\bm{\Lambda}\neq 0$, depends on the input states $\{\rho ^{(i)}\}$
and the noise operations $\bm{\chi}^{(\text{i})}$ and
$\bm{\chi}^{(\text{f})}$. It
may happen that the $\bm{\Lambda}$ matrix becomes ill-conditioned \cite%
{meyer} for a specific set of input states (for some given noise operations $%
\bm{\chi}^{(\text{i})}$ and $\bm{\chi}^{(\text{f})}$). In such cases, even
small errors (whether operational, stochastic, or round-off) in estimation of $%
\bm{\chi}^{(\text{T})}$ can be amplified dramatically after multiplication by $%
\bm{\Lambda}^{-1}$. This in turn may render the estimation of
$\bm{\chi}$ (the sought-for unknown map $\mathcal{E}$) completely
unreliable. To minimize the statistical errors, the input states
should be chosen such that $\det \bm{\Lambda}$ is as far from zero
as possible. Therefore, the optimal input states
$\{\rho_{\text{opt}}^i\}$ [optimal in the sense of minimizing
statistical errors] for given $\bm{\chi}^{(\text{i})}$ and
$\bm{\chi}^{(\text{f})}$ are obtained via maximizing $\det
\bm{\Lambda}$. A similar \textit{faithfulness} measure has already
been used in Refs.~\cite{altepeter-aapt,aapt}. In
Appendix~\ref{id-inp-dcqd}, we derive the optimal input states for
the case of
the
ideal DCQD
scheme.

\section{Process estimation with specific noisy devices}
\label{models}

In the following we describe several examples which describe relevant physical noise models.

\subsection{Depolarizing channels: Correlated noise}

An important and practically relevant example is the situation in which $%
\mathcal{E}^{(\text{i})}$ and $\mathcal{E}^{(\text{f})}$ both are
two-qubit (hence correlated) depolarizing channels $\mathcal{D}^{[2]}$ \cite{depol,2qbit-depolarizing}
\begin{eqnarray*}
\rho^{(i)}\overset{\mathcal{D}^{[2]}_{\varepsilon}}{\rightarrow} \frac{%
1-\varepsilon}{4}\openone\otimes\openone + \varepsilon\rho^{(i)}, \\
P^{jj}\overset{\mathcal{D}^{[2]}_{\varepsilon^{\prime }}}{\rightarrow} \frac{%
1-\varepsilon^{\prime }}{4}\openone\otimes \openone + \varepsilon^{\prime
jj},
\end{eqnarray*}
where $\varepsilon$ and $\varepsilon'$ could be independent of each other or correlated (e.g., $\varepsilon=\varepsilon'$).
These errors result in the following noisy data processing of the measurement
results of DCQD:
\begin{widetext}
\begin{eqnarray}
\text{Tr}[\mathcal{E}(\rho^{(i)})P^{jj}] \rightarrow
\ignore{
 \text{Tr}\left[
\mathcal{E}\left(\frac{1-\varepsilon}{4}\openone\otimes\openone +\varepsilon
\rho^{(i)} \right) ( \frac{1-\varepsilon^{\prime }}{4}\openone\otimes %
\openone+\varepsilon^{\prime jj})\right]  \notag \\
& =&
}
 \frac{(1-\varepsilon)(1-\varepsilon^{\prime })}{16} \text{Tr}[\mathcal{E%
}(\openone) \otimes\openone] + \frac{\varepsilon^{\prime }(1-\varepsilon)}{4}
\text{Tr}[\mathcal{E} (\openone)\otimes \openone P^{jj}] + \frac{%
\varepsilon(1-\varepsilon^{\prime })}{4} \text{Tr}[\mathcal{E}(\rho^{(i)})]
+ \varepsilon\varepsilon^{\prime }\text{Tr}[\mathcal{E}(\rho^{(i)})P^{jj}].
\label{corr-noise}
\end{eqnarray}
\end{widetext}
For the Hamiltonian identification task \cite{ham-id,ham-id2}, $\mathcal{E}%
(\rho)=e^{-iHt}\rho e^{iHt}$ (which is unital: $\mathcal{E}(\openone)=\openone$,
and trace-preserving: $\text{Tr}[\mathcal{E}(\rho)]=1$), we obtain
\begin{eqnarray}
\text{Tr}[\mathcal{E}(\rho^{(i)})P^{jj}] \rightarrow
\varepsilon\varepsilon^{\prime }\text{Tr}[\mathcal{E}(\rho^{(i)})P^{jj}]+
(1-\varepsilon\varepsilon^{\prime })/4.  \label{rules}
\end{eqnarray}
This relation provides a simple connection between the ideal and
the noisy data processing rules. Another feature of Eq.~(\ref{rules}) is
that it is valid irrespective of the values of $\varepsilon$ and
$\varepsilon'$ ($\neq0$). This implies that, whether $\varepsilon$
and $\varepsilon'$ are in the range which makes the noisy
preparation/BSM separable or not \cite{separability,noisy}, the
simplicity and applicability of (the modified) DCQD remain intact.
In other words, entanglement is not an imperative in the DCQD
algorithm.

A generalization of this noise model is the case in which the preparations are
modified based on a generalized two-qubit depolarizing channels \cite{2qbit-depolarizing}:
\begin{eqnarray*}
\rho^{(i)}\overset{\widetilde{\mathcal{D}}^{[2]}_{\varepsilon}}{\rightarrow} \frac{%
1-\varepsilon}{4}\openone\otimes\openone + \varepsilon U\rho^{(i)}U^{\dag},
\end{eqnarray*}
in which $U$ is an already known two-qubit unitary operator. To simplify the
following discussion we assume that BSMs are noiseless ($\mathcal{E}^{(\text{%
f})}=\mathbb{I}$). Finding the explicit form of $\bm{\chi}^{(i)}$ is
straightforward. We use the form
\begin{eqnarray*}
& \rho=\frac{1}{4}(\openone\otimes\openone+{
\sum'}_{m,n=0}^3 r_{mn}\sigma_m\otimes\sigma_n),
\end{eqnarray*}
where $\sum'$ denotes the constrained
summation in which the case $(m,n)=(0,0)$ has been
excluded. Using the identity $\sigma_k\sigma_l\sigma_k=(-1)^{1-\delta_{kl}}\sigma_{l}$,
we have: $\frac{1}{4}\sum_{ab=0}^3\sigma_a\otimes
\sigma_b\rho\sigma_a\otimes\sigma_b=\rho + \frac{3}{4}\openone\otimes\openone
$, or equivalently: $\sum_{ab=0}^3p_{ab}\sigma_a\otimes\sigma_b\rho\sigma_a
\otimes\sigma_b=\openone\otimes\openone$, where $p_{ab}=1/3$ except for
$p_{00}=-1$. In addition, we expand $U$ in the $\{\sigma_m\otimes\sigma_n%
\}_{mn=0}^3$ basis: $U=\sum_{mn}a_{mn}\sigma_m\otimes\sigma_n$.
Altogether, these relations yield
\begin{eqnarray*}
\widetilde{\mathcal{D}}^{[2]}_{\varepsilon}(\rho)
&=&\frac{1-\varepsilon}{4}\sum_{mn}p_{mn} \sigma_m\otimes\sigma_n\rho\sigma_m\otimes\sigma_n\\
&&+\varepsilon\sum_{mn,m'n'}a_{mn}\bar{a}_{m'n'}
\sigma_m\otimes\sigma_n\rho \sigma_{m'}\otimes\sigma_{n'}.
\end{eqnarray*}
Hence, we obtain: $\chi^{(\text{i})}_{mnmn}=p_{mn}(1-\varepsilon)/4+
\varepsilon|a_{mn}|^2$ (the diagonal elements) and
$\chi^{(\text{i})}_{mnm'n'}=\varepsilon a_{mn}\bar{a}_{m'n'}$ for
$(m,n)\neq(m',n')$ (the off-diagonal elements). In a compact form, the effect
of this noise channel can be expressed as follows:
\begin{eqnarray}
&&\hskip -2mm\text{Tr}[\mathcal{E}(\rho^{(i)})P^{jj}] \rightarrow\nonumber\\
\ignore{
\text{Tr}\left[
\mathcal{E}\left(\frac{1-\varepsilon}{4}\openone\otimes\openone +\varepsilon
U\rho^{(i)}U^{\dag} \right) ( \frac{1-\varepsilon^{\prime }}{4}\openone%
\otimes \openone+\varepsilon^{\prime jj})\right]  \notag \\
& =&
}
&&~\frac{(1-\varepsilon)(1-\varepsilon^{\prime })}{16} \text{Tr}[\mathcal{E%
}(\openone) \otimes\openone] + \frac{\varepsilon^{\prime }(1-\varepsilon)}{4}
\text{Tr}[\mathcal{E}(\openone) \otimes \openone P^{jj}] \nonumber\\
&&~+ \frac{%
\varepsilon(1-\varepsilon^{\prime })}{4} \text{Tr}[\mathcal{E}%
(U\rho^{(i)}U^{\dag})] + \varepsilon\varepsilon^{\prime }\text{Tr}[\mathcal{E}(U\rho^{(i)}U^{%
\dag})P^{jj}].
\label{corr-noise-2}
\end{eqnarray}
Under trace-preserving and unitality conditions, the final data
processing is thus modified as follows:
\begin{eqnarray}
\text{Tr}[\mathcal{E}(\rho^{(i)})P^{jj}] \rightarrow \varepsilon
\varepsilon^{\prime }\text{Tr}[\mathcal{E}(U\rho^{(i)}U^{\dag})P^{jj}]+
(1-\varepsilon\varepsilon^{\prime })/4.  \label{rules-2}
\end{eqnarray}
Although this is not as simple as Eq.~(\ref{rules}), it yet retains a considerable simplicity.

\subsection{Depolarizing channels: Uncorrelated noise}

We assume that the input states and our measurements are diluted by
depolarizing channels \cite{noisy,separability} acting \textit{separately}
on the principal and ancilla qubits, i.e., $\mathcal{D}\otimes\mathcal{D}$,
where $\mathcal{D}$ acts on a general single-qubit state $\rho$ as
follows: $\mathcal{D}_{\varepsilon}(\rho)=\frac{1-\varepsilon}{2}\openone%
+ \varepsilon\rho$, or equivalently: $\mathcal{D}_{\varepsilon}(\rho)=%
\sum_{j=0}^3p_j\sigma_j\rho\sigma_j$, where $p_0=(1+3\varepsilon)/4$ and $%
p_1=p_2=p_3=(1-\varepsilon)/4$, and positivity and complete-positivity of
$\mathcal{D}_{\varepsilon}$ require $-1/3\leqslant\varepsilon\leqslant 1$ \cite{dep-cp}.

As a special case we specialize on the characterization of the
diagonal elements $\chi_{kk}$. This is particularly important in
Hamiltonian identification tasks \cite{ham-id,ham-id2}. It can be
easily seen that for Bell-states $P^{kk}$
we obtain
\begin{eqnarray*}
P^{kk}\overset{\mathcal{D}_{\varepsilon} \otimes\mathcal{D}_{\varepsilon}}{%
\longrightarrow}\frac{1-\varepsilon^2}{4}\openone \otimes \openone+
\varepsilon^2 P^{kk}.
\end{eqnarray*}
Thus, to estimate $\chi_{kk}$, the necessary data processing is
modified as in Eqs.~(\ref{corr-noise}) and (\ref{rules}) by replacing: $%
\varepsilon\varepsilon'\rightarrow (\varepsilon\varepsilon')^2$ and
$i\rightarrow 0$ (recall that $\rho^{(0)}=|\Phi^+\rangle\langle
\Phi^+|$). Here we have assumed that the input (measurement)
depolarizing parameter is $\varepsilon$ ($\varepsilon^{\prime }$).
This result implies that to estimate the diagonal elements
$\chi_{kk}$, whether under correlated noise or uncorrelated noise,
the DCQD scheme is robust and classical data processing is modified
in a simple fashion. This has
immediate applications to the task of Hamiltonian identification \cite%
{ham-id}.

\subsection{Generalized depolarizing channels}

Here, we assume that the input states and/or measurements are
diluted such that they effectively lead to (known) Bell-diagonal
input states and/or Bell-diagonal measurements.
Thus we obtain
\begin{eqnarray*}
&\rho^{(i)}\overset{\mathcal{B}_{\varepsilon}}{\rightarrow} \sum_{i^{\prime
}=0}^3\varepsilon_{ii^{\prime }}\rho^{(i^{\prime })}, \\
& P^{jj}\overset{\mathcal{B}_{\varepsilon^{\prime }}}{\rightarrow}
\sum_{j^{\prime }=0}^3 \varepsilon^{\prime }_{jj^{\prime }} P^{j^{\prime
}j^{\prime }}.
\end{eqnarray*}
This noise results in the following noisy data processing of the measurement
results of DCQD:
\begin{eqnarray}
&\text{Tr}[\mathcal{E}(\rho^{(i)})P^{jj}] \rightarrow
\ignore{
 \text{Tr}\left[
\mathcal{E}\left(\sum_{i^{\prime }=0}^3\varepsilon_{ii^{\prime
}}\rho^{(i^{\prime })}\right) ( \sum_{j^{\prime }=0}^3 \varepsilon^{\prime
}_{jj^{\prime }} P^{j^{\prime }j^{\prime }})\right]  \notag \\
& =&
}
\sum_{i^{\prime }j^{\prime }}\varepsilon_{ii^{\prime
}}\varepsilon^{\prime }_{jj^{\prime }}\text{Tr}[\mathcal{E}(\rho^{(i^{\prime
})})P^{j^{\prime }j^{\prime }}].
\label{noise}
\end{eqnarray}
That is, every measurement result of the new setting is a linear combination
of the ideal results. If we define the vector $|\mathbf{p})=(\mathbf{p}_{ij})^T$, where $\mathbf{p}_{ij}=\text{Tr}[\mathcal{E}(\rho^{(i)})P^{jj}]$, namely
\begin{eqnarray*}
& \hskip -2mm|\mathbf{p})=\bigl(\text{Tr}[\mathcal{E}(\rho^{(0)})P^{00}],\text{Tr}[\mathcal{E}(\rho^{(0)})P^{11}],\ldots, \text{Tr}[\mathcal{E}(\rho^{(3)})P^{33}]\bigr)^T,
\end{eqnarray*}
and the matrix $\mathbf{A}_{ij,i^{\prime
}j^{\prime }}=\varepsilon_{ii^{\prime }}\varepsilon^{\prime }_{jj^{\prime
}}$, then (\ref{noise}) can be written as the following linear matrix
transformation (see Appendix~\ref{app-1}):
\begin{eqnarray}
&|\mathbf{p})\rightarrow \mathbf{A}|\mathbf{p}).
\end{eqnarray}
If we arrange the output elements as in Table~\ref{dcqd-tab}, we will have
\begin{eqnarray}
&| \widetilde{\mathbf{p}}) =\mathbf{C}|\mathbf{p}),
\label{big-eq-1}
\end{eqnarray}
where $\mathbf{C}$ is the (constant) coefficient matrix, hence $|\widetilde{\mathbf{p}}) \rightarrow \mathbf{A} \mathbf{C}^{-1}|\widetilde{\mathbf{p}})$.


\section{Summary}
\label{summary}

We have provided a scheme for utilizing auxiliary quantum
correlations to perform process estimation tasks with faulty quantum
operations. We have demonstrated our approach via generalizing the
ideal scheme of Direct Characterization of Quantum Dynamics (DCQD)
where the required preparations and measurements could be noisy. It has been shown that when the systematic faulty operations are of the
form of depolarizing channels, the overhead data processing is
fairly simple. Moreover, these examples have revealed that for the
DCQD scheme, entanglement is secondary. This, in turn, broadens the
range of applicability of our scheme to quantum systems with certain
controllable classical correlations of their subsystems. Therefore,
our proposed method may have near-term applications to a variety of
realistic quantum systems/devices with the current
state-of-technology, such as trapped ions, liquid-state NMR, optical
lattices, and entangled pairs of photons.

We thank Natural Sciences and Engineering Research Council of Canada
(NSERC), Faculty of Arts and Sciences of Harvard University, Army Research Office (ARO) [project W911NF-07-1-0304], Mathematics of Information Technology and Complex Systems (MITACS), Pacific Institute for Mathemaical Sciences (PIMS), and the USC Center for Quantum Information Science and Technology for funding.

\begin{widetext}
\appendix

\section{Explicit form of Eq.~(\ref{big-eq-1})}
\label{app-1} Table~\ref{dcqd-tab} suggests that if, instead of the
conventional BSMs, we consider the expression
\begin{eqnarray}
\left(
\begin{smallmatrix}
\mathrm{Tr}[P^{00}\mathcal{E}(\rho ^{(0)})] \\
\mathrm{Tr}[P^{11}\mathcal{E}(\rho ^{(0)})] \\
\mathrm{Tr}[P^{22}\mathcal{E}(\rho ^{(0)})] \\
\mathrm{Tr}[P^{33}\mathcal{E}(\rho ^{(0)})] \\
\mathrm{Tr}[P^{00}\mathcal{E}(\rho ^{(1)})]+\mathrm{Tr}[P^{33}\mathcal{E}%
(\rho ^{(1)})] \\
\mathrm{Tr}[P^{11}\mathcal{E}(\rho ^{(1)})]+\mathrm{Tr}[P^{22}\mathcal{E}%
(\rho ^{(1)})] \\
\mathrm{Tr}[P^{00}\mathcal{E}(\rho ^{(1)})]-\mathrm{Tr}[P^{33}\mathcal{E}%
(\rho ^{(1)})] \\
\mathrm{Tr}[P^{11}\mathcal{E}(\rho ^{(1)})]-\mathrm{Tr}[P^{22}\mathcal{E}%
(\rho ^{(1)})] \\
\mathrm{Tr}[P^{00}\mathcal{E}(\rho ^{(2)})]+\mathrm{Tr}[P^{11}\mathcal{E}%
(\rho ^{(2)})] \\
\mathrm{Tr}[P^{22}\mathcal{E}(\rho ^{(2)})]+\mathrm{Tr}[P^{33}\mathcal{E}%
(\rho ^{(2)})] \\
\mathrm{Tr}[P^{00}\mathcal{E}(\rho ^{(2)})]-\mathrm{Tr}[P^{11}\mathcal{E}%
(\rho ^{(2)})] \\
\mathrm{Tr}[P^{33}\mathcal{E}(\rho ^{(2)})]-\mathrm{Tr}[P^{22}\mathcal{E}%
(\rho ^{(2)})] \\
\mathrm{Tr}[P^{00}\mathcal{E}(\rho ^{(3)})]+\mathrm{Tr}[P^{22}\mathcal{E}%
(\rho ^{(3)})] \\
\mathrm{Tr}[P^{11}\mathcal{E}(\rho ^{(3)})]+\mathrm{Tr}[P^{33}\mathcal{E}%
(\rho ^{(3)})] \\
\mathrm{Tr}[P^{00}\mathcal{E}(\rho ^{(3)})]-\mathrm{Tr}[P^{22}\mathcal{E}%
(\rho ^{(3)})] \\
\mathrm{Tr}[P^{33}\mathcal{E}(\rho ^{(3)})]-\mathrm{Tr}[P^{11}\mathcal{E}%
(\rho ^{(3)})]%
\end{smallmatrix}
\right) = \left(%
\begin{smallmatrix}
1 &  &  &  &  &  &  &  &  &  &  &  &  &  &  &  \\
& 1 &  &  &  &  &  &  &  &  &  &  &  &  &  &  \\
&  & 1 &  &  &  &  &  &  &  &  &  &  &  &  &  \\
&  &  & 1 &  &  &  &  &  &  &  &  &  &  &  &  \\
&  &  &  & 1 &  &  & 1 &  &  &  &  &  &  &  &  \\
&  &  &  &  & 1 & 1 &  &  &  &  &  &  &  &  &  \\
&  &  &  & 1 &  &  & -1 &  &  &  &  &  &  &  &  \\
&  &  &  &  & 1 & -1 &  &  &  &  &  &  &  &  &  \\
&  &  &  &  &  &  &  & 1 & 1 &  &  &  &  &  &  \\
&  &  &  &  &  &  &  &  &  & 1 & 1 &  &  &  &  \\
&  &  &  &  &  &  &  & 1 & -1 &  &  &  &  &  &  \\
&  &  &  &  &  &  &  &  &  & -1 & 1 &  &  &  &  \\
&  &  &  &  &  &  &  &  &  &  &  & 1 &  & 1 &  \\
&  &  &  &  &  &  &  &  &  &  &  &  & 1 &  & 1 \\
&  &  &  &  &  &  &  &  &  &  &  & 1 &  & -1 &  \\
&  &  &  &  &  &  &  &  &  &  &  &  & -1 &  & 1%
\end{smallmatrix}%
\right)\left(%
\begin{smallmatrix}
\mathrm{Tr}[P^{00}\mathcal{E}(\rho ^{(0)})] \\
\mathrm{Tr}[P^{11}\mathcal{E}(\rho ^{(0)})] \\
\mathrm{Tr}[P^{22}\mathcal{E}(\rho ^{(0)})] \\
\mathrm{Tr}[P^{33}\mathcal{E}(\rho ^{(0)})] \\
\mathrm{Tr}[P^{00}\mathcal{E}(\rho^{(1)})] \\
\mathrm{Tr}[P^{11}\mathcal{E}(\rho^{(1)})] \\
\mathrm{Tr}[P^{22}\mathcal{E}(\rho^{(1)})] \\
\mathrm{Tr}[P^{33}\mathcal{E}(\rho^{(1)})] \\
\mathrm{Tr}[P^{00}\mathcal{E}(\rho^{(2)})] \\
\mathrm{Tr}[P^{11}\mathcal{E}(\rho^{(2)})] \\
\mathrm{Tr}[P^{22}\mathcal{E}(\rho^{(2)})] \\
\mathrm{Tr}[P^{33}\mathcal{E}(\rho^{(2)})] \\
\mathrm{Tr}[P^{00}\mathcal{E}(\rho^{(3)})] \\
\mathrm{Tr}[P^{11}\mathcal{E}(\rho^{(3)})] \\
\mathrm{Tr}[P^{22}\mathcal{E}(\rho^{(3)})] \\
\mathrm{Tr}[P^{33}\mathcal{E}(\rho^{(3)})]%
\end{smallmatrix}%
\right),
\label{parametrization}
\end{eqnarray}
the results of the measurements can be related to the $\chi_{mn}$
elements in a more straightforward fashion. Here, the constant
coefficient matrix ($\mathbf{C}$) relates the results of the new
BSMs ($\widetilde{\mathbf{p}}$) to those of the conventional BSMs
($\mathbf{p}$).

\section{Optimal input states for the ideal (noiseless) DCQD}
\label{id-inp-dcqd}

\begin{figure}[tp]
\includegraphics[width=5.5cm,height=4cm]{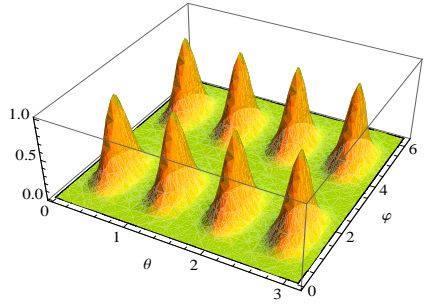}
\caption{The value of $|\det\bm{\Lambda}|$ vs $\theta$ and
$\varphi$. Here, the coefficient matrix
$\bm{\Lambda}(\theta,\varphi)$ relates the experimental outcomes to
the unknown elements of the superoperator $|\bm{\chi}^{(\text{T})})
=\mathbf{\Lambda}|\bm{\chi})$. The input states for the standard
DCQD, as defined in Table~\ref{dcqd-tab}, e.g.,
$|\Phi^{+}_{\alpha}\rangle=\alpha|00\rangle+\beta|11\rangle$, are
parameterized as $\alpha=\cos\theta$ and $%
\beta=e^{i\varphi}\sin{\theta}$ ($\varphi\neq k\pi$,
$k\in\mathbb{Z}$). The optimal input states are associated to those
parameters for which $|\det\bm{\Lambda}|$ has its maximal value 1,
leading to a minimal statistical error.} \label{maxima}
\end{figure}

Here, we find the optimal input states for the ideal DCQD. The idea
is to choose the input states such that the (linear) inversion on
the experimental data (to read out $\bm{\chi}$ matrix elements) can
be performed reliably. That is, the goal should be to make the
coefficient matrix as far from singular matrices as possible.
Maximizing the determinant of this matrix is a sufficient condition
to guarantee its reliable invertibility, hence in turn, minimal
error propagation.

The data obtained from the measurements (BSMs) are $\text{Tr}\left[\mathcal{E}%
(\rho^{(i)})P^{jj} \right]$, where $\{\rho^{(i)}\}$ corresponds to the first
column of Table~\ref{dcqd-tab}, respectively, for $i=0,1,2,3$. We parameterize the input states by $\alpha=\cos\theta$ and $%
\beta=e^{i\varphi}\sin{\theta}$ ($\varphi\neq k\pi$,
$k\in\mathbb{Z}$). Using relation (\ref{parametrization}), one can
express Eq.~(\ref{vector-eq}), for the standard DCQD \cite{dcqd1},
as the following:
\begin{eqnarray*}
\left(
\begin{smallmatrix}
\mathrm{Tr}[P^{00}\mathcal{E}(\rho ^{(1)})] \\
\mathrm{Tr}[P^{11}\mathcal{E}(\rho ^{(1)})] \\
\mathrm{Tr}[P^{22}\mathcal{E}(\rho ^{(1)})] \\
\mathrm{Tr}[P^{33}\mathcal{E}(\rho ^{(1)})] \\
\mathrm{Tr}[P^{00}\mathcal{E}(\rho ^{(2)})]+\mathrm{Tr}[P^{33}\mathcal{E}%
(\rho ^{(2)})] \\
\mathrm{Tr}[P^{11}\mathcal{E}(\rho ^{(2)})]+\mathrm{Tr}[P^{22}\mathcal{E}%
(\rho ^{(2)})] \\
\mathrm{Tr}[P^{00}\mathcal{E}(\rho ^{(2)})]-\mathrm{Tr}[P^{33}\mathcal{E}%
(\rho ^{(2)})] \\
\mathrm{Tr}[P^{11}\mathcal{E}(\rho ^{(2)})]-\mathrm{Tr}[P^{22}\mathcal{E}%
(\rho ^{(2)})] \\
\mathrm{Tr}[P^{00}\mathcal{E}(\rho ^{(3)})]+\mathrm{Tr}[P^{11}\mathcal{E}%
(\rho ^{(3)})] \\
\mathrm{Tr}[P^{22}\mathcal{E}(\rho ^{(3)})]+\mathrm{Tr}[P^{33}\mathcal{E}%
(\rho ^{(3)})] \\
\mathrm{Tr}[P^{00}\mathcal{E}(\rho ^{(3)})]-\mathrm{Tr}[P^{11}\mathcal{E}%
(\rho ^{(3)})] \\
\mathrm{Tr}[P^{33}\mathcal{E}(\rho ^{(3)})]-\mathrm{Tr}[P^{22}\mathcal{E}%
(\rho ^{(3)})] \\
\mathrm{Tr}[P^{00}\mathcal{E}(\rho ^{(4)})]+\mathrm{Tr}[P^{22}\mathcal{E}%
(\rho ^{(4)})] \\
\mathrm{Tr}[P^{11}\mathcal{E}(\rho ^{(4)})]+\mathrm{Tr}[P^{33}\mathcal{E}%
(\rho ^{(4)})] \\
\mathrm{Tr}[P^{00}\mathcal{E}(\rho ^{(4)})]-\mathrm{Tr}[P^{22}\mathcal{E}%
(\rho ^{(4)})] \\
\mathrm{Tr}[P^{33}\mathcal{E}(\rho ^{(4)})]-\mathrm{Tr}[P^{11}\mathcal{E}%
(\rho ^{(4)})]%
\end{smallmatrix}
\right) = \left(
\begin{smallmatrix}
1 & 0 & 0 & 0 & 0 & 0 & 0 & 0 & 0 & 0 & 0 & 0 & 0 & 0 & 0 & 0 \\
0 & 0 & 0 & 0 & 0 & 1 & 0 & 0 & 0 & 0 & 0 & 0 & 0 & 0 & 0 & 0 \\
0 & 0 & 0 & 0 & 0 & 0 & 0 & 0 & 0 & 0 & 1 & 0 & 0 & 0 & 0 & 0 \\
0 & 0 & 0 & 0 & 0 & 0 & 0 & 0 & 0 & 0 & 0 & 0 & 0 & 0 & 0 & 1 \\
1 & 0 & 0 & x & 0 & 0 & 0 & 0 & 0 & 0 & 0 & 0 & x & 0 & 0 & 1 \\
0 & 0 & 0 & 0 & 0 & 1 & -i x & 0 & 0 & i x & 1 & 0 & 0 & 0 & 0 & 0 \\
z & 0 & 0 & i y & 0 & 0 & 0 & 0 & 0 & 0 & 0 & 0 & -i y & 0 & 0 & -z \\
0 & 0 & 0 & 0 & 0 & z & y & 0 & 0 & y & -z & 0 & 0 & 0 & 0 & 0 \\
1 & x & 0 & 0 & x & 1 & 0 & 0 & 0 & 0 & 0 & 0 & 0 & 0 & 0 & 0 \\
0 & 0 & 0 & 0 & 0 & 0 & 0 & 0 & 0 & 0 & 1 & -i x & 0 & 0 & i x & 1 \\
z & i y & 0 & 0 & -i y & -z & 0 & 0 & 0 & 0 & 0 & 0 & 0 & 0 & 0 & 0 \\
0 & 0 & 0 & 0 & 0 & 0 & 0 & 0 & 0 & 0 & z & y & 0 & 0 & y & -z \\
0 & 0 & 0 & 0 & 0 & 1 & 0 & -i x & 0 & 0 & 0 & 0 & 0 & i x & 0 & 1 \\
1 & 0 & -x & 0 & 0 & 0 & 0 & 0 & -x & 0 & 1 & 0 & 0 & 0 & 0 & 0 \\
0 & 0 & 0 & 0 & 0 & -z & 0 & -y & 0 & 0 & 0 & 0 & 0 & -y & 0 & z \\
-z & 0 & i y & 0 & 0 & 0 & 0 & 0 & -i y & 0 & z & 0 & 0 & 0 & 0 & 0%
\end{smallmatrix}
\right) \left(%
\begin{smallmatrix}
\chi _{00} \\
\chi _{01} \\
\chi _{02} \\
\chi _{03} \\
\chi _{10} \\
\chi _{11} \\
\chi _{12} \\
\chi _{13} \\
\chi _{20} \\
\chi _{21} \\
\chi _{22} \\
\chi _{23} \\
\chi _{30} \\
\chi _{31} \\
\chi _{32} \\
\chi _{33}%
\end{smallmatrix}%
\right),
\end{eqnarray*}
where in the coefficient matrix $\bm{\Lambda}(\theta,\varphi)$ we have $%
x=\cos 2\theta$, $y=\sin 2 \theta \sin \varphi$, and $z= \sin 2 \theta\cos
\varphi $. The determinant of this matrix is obtained as
\begin{eqnarray*}
&|\det\bm{\Lambda}|=\sin^6 4\theta \sin^6\varphi,
\end{eqnarray*}
which attains its maximum value $1$ at $\left(\theta=\pi/8+k\pi/4,
\varphi=\pi/2+k^{\prime }\pi\right)$, $\forall k,k^{\prime }\in\mathbb{Z}$ --- Fig.~\ref{maxima}.
Therefore, the optimal input states $\{\rho^{(i)}\}$ for the standard DCQD
are as in Table~\ref{dcqd-tab} in which $\mu$ and $\nu$ are either of the
pairs calculated from the above maximal set of $\theta$ and $%
\varphi $. A simple calculation shows that the amount of entanglement
(exactly speaking, concurrence \cite{concurrence}) of the optimal
non-maximally entangled input states is $1/\sqrt{2}$ (independent of $%
\varphi $).

\twocolumngrid
\end{widetext}


\end{document}